\documentclass[apj,numberedappendix,twocolumn]{openjournal}

\usepackage{orcidlink}
\usepackage{tikz}
\usepackage{multirow}
\usepackage{colortbl}
\usepackage{float}

\usepackage{amsfonts,amsmath,graphicx,natbib,apjfonts,subfigure,xcolor,gensymb,longtable,appendix,hyperref,placeins}

\definecolor{xlinkcolor}{cmyk}{1,1,0,0}
\hypersetup{
    pdfnewwindow=true,      
    colorlinks=true,    
    linkcolor=xlinkcolor,     
    citecolor=xlinkcolor,     
    filecolor=xlinkcolor,  
    urlcolor=xlinkcolor
}

\graphicspath{{./}{figures/}}

\begin{document}

\title{Implications of the X-ray radiation from the hydrogen-poor SLSN\,2018bsz in the context of shock-driven emission}

\author{Sarah Pedrami$^{1,2}$\orcidlink{0009-0001-3541-6853}}
\author{Raffaella Margutti$^{1,2,3}$\orcidlink{0000-0003-4768-7586}}
\author{Eli Wiston$^{1,2}$\orcidlink{0009-0002-4843-2913}}
\author{Maria~R. Drout$^4$\orcidlink{0000-0001-7081-0082}}
\author{Daisy Matthews$^5$\orcidlink{0000-0002-4513-3849}}
\author{Laura Chomiuk$^6$\orcidlink{0000-0002-8400-3705}}
\author{Ryan Chornock$^{1,2}$\orcidlink{0000-0002-7706-5668}}
\author{Wynn~V. Jacobson-Gal\'an$^{7,\star}$\orcidlink{0000-0002-3934-2644}}
\author{Joseph~R. Farah$^{1,8}$\orcidlink{0000-0003-4914-5625}}
\author{Deanne~L. Coppejans$^9$\orcidlink{0000-0001-5126-6237}}
\author{Brian~D. Metzger$^{10,11}$\orcidlink{0000-0002-4670-7509}}
\author{Dan Milisavljevic$^{12}$\orcidlink{0000-0002-0763-3885}}
\author{Itai Sfaradi$^{1,2}$\orcidlink{0000-0003-0466-3779}}
\author{Edo Berger$^{13}$\orcidlink{0000-0002-9392-9681}}

\affiliation{$^1$ Department of Astronomy, University of California, Berkeley, CA 94720-3411, USA\\
$^2$ Berkeley Center for Multi-messenger Research on Astrophysical Transients and Outreach (Multi-RAPTOR), University of California, Berkeley, CA 94720-3411, USA\\
$^3$ Department of Physics, University of California, 366 Physics North MC 7300,
Berkeley, CA 94720, USA\\
$^4$ David A. Dunlap Department of Astronomy and Astrophysics, University of Toronto \\ 50 St. George Street, Toronto, Ontario, M5S 3H4 Canada\\
$^5$ Department of Physics, Illinois Institute of Technology, Chicago, IL 60616, USA\\
$^6$ Center for Data Intensive and Time Domain Astronomy, Department of Physics and Astronomy, Michigan State University, East Lansing, MI 48824, USA\\
$^7$ Cahill Center for Astrophysics, California Institute of Technology, MC 249-17, 1216 E California Boulevard, Pasadena, CA, 91125, USA\\
$^8$ Miller Institute for Basic Research in Science, 206B Stanley Hall, Berkeley, CA 94720, USA\\
$^9$ Department of Physics, University of Warwick, Coventry CV4 7AL, UK\\
$^{10}$ Department of Physics and Columbia Astrophysics Laboratory, Columbia University, New York, NY 10027, USA\\
$^{11}$ Center for Computational Astrophysics, Flatiron Institute, 162 5th Avenue, New York, NY 10010, USA\\
$^{12}$ Department of Physics and Astronomy, Purdue University, 525 Northwestern Avenue, West Lafayette, IN 47907, USA\\
$^{13}$ Center for Astrophysics | Harvard \& Smithsonian, 60 Garden Street, Cambridge, MA 02138-1516, USA\\
$^\star$ NASA Hubble Fellow}

\begin{abstract}
At a distance of 111\,Mpc, SN\,2018bsz is one of the nearest hydrogen-poor superluminous supernovae (SLSNe-I) to date, distinguished by (i) a long pre-peak emission plateau, (ii) late-time emergence of hydrogen features at $\delta t = 75\textrm{--}105$\,d, and (iii) NIR/optical light curve modulations at $\delta t < 100$\,d. Here, we analyze X-ray observations from our multi-epoch Chandra X-ray Observatory (CXO) and \textit{Swift}-XRT campaigns spanning $\delta t = 87\textrm{--}303$\,d after explosion. Reaching $L_{\rm x} \approx 2 \times 10^{40}\,\rm{erg\,s^{-1}}$, the X-ray luminosity at the location of SN\,2018bsz is significantly larger than normal Type Ibc and Type II SNe at comparable epochs and rivals the lower end of the Type IIn population, with $\approx 3\sigma$ variability indicating a transient-dominated component. To move beyond single-shock scalings that neglect time-dependent opacity, we develop a semi-analytic framework modeling both forward and reverse shock X-ray emission in a wind-like circumstellar medium (CSM), incorporating energy-dependent photoelectric absorption from the unshocked CSM and cold dense shell (CDS). As CDS absorption diminishes, reverse shock emission emerges and can dominate the observed flux. Applying this model to our CXO and \textit{Swift} dataset, under the assumption that the X-ray emission originates from the SN, yields an effective progenitor mass-loss rate $\dot{M} = (10^{-3}\textrm{--}10^{-2})\,\rm M_\odot\,yr^{-1}$ (for $v_{\rm w} = 1000\,\rm km\,s^{-1}$). Late-time X-ray observations at $\delta t > 2000$\,d will directly test and potentially rule out this scenario, and will ultimately prove if the X-ray emission is powered by the SN.
\end{abstract}


\maketitle

\section{Introduction}
\label{Sec:intro}
Superluminous supernovae (SLSNe) are a type of transient characterized by peak optical luminosities of $L_{\rm pk} \gtrsim 10^{43}\,\rm{erg\,s^{-1}}$, corresponding to $\sim10\textrm{--}100$ times that of standard supernovae (SN; \citealt{Gal-Yam12, Chomiuk11, Quimby11}). SLSNe are observationally divided into hydrogen-poor (SLSN-I) and hydrogen-rich (SLSN-II) subtypes (e.g., \citealt{galyam19, moriya26} for a review). Hydrogen is commonly present in the spectra of some SLSNe-II in the form of narrow multi-component emission lines (e.g., \citealt{smith07}). These features are indicative of strong shock interaction between the SN ejecta and surrounding material in the circumstellar medium (CSM). In this case, the ejecta--CSM interaction can convert a large fraction of the ejecta kinetic energy into radiation, powering the observed emission \citep{chev82, chevalier17}. SLSNe-II with these features are classified as ``SLSNe-IIn'' for their similarity to fainter SNe-IIn.

SLSNe-I, in contrast, exhibit spectral features distinct from other SN types. Their early-time spectra are typically characterized by a blue continuum and broad absorption features attributed to O\,\textsc{ii} (e.g., \citealt{quimby18}). The power source of these events remains under debate. One possibility is interaction between the ejecta and a dense, H-poor CSM, but clear spectroscopic evidence of this mechanism is missing \citep{chevalier11, chatz13}. Moreover, a small number of SLSNe-I have shown evidence for the late emergence of hydrogen features, possibly indicating interaction with a hydrogen-rich CSM (see e.g., \citealt{Yan15,yan17a}). Alternative explanations for the large SLSNe-I luminosities involve energy injection from a spin-down magnetar central engine
(e.g., \citealt{kasen10,Dessart19,chenk16,Moriya16}). In this case, the deposited energy from the central engine heats the ejecta, powering the high luminosities. 

Both CSM interaction and central engines are known sources of X-ray emission. In the shock CSM interaction scenario, the conversion of ejecta kinetic energy into radiation generates thermal X-rays due to bremsstrahlung emission \citep{Fransson96, chevalier17}, as observed for strongly-interacting SNe like SN\,2014C \citep{Brethauer22,Thomas22}. In the central engine scenario, X-ray emission is typically associated with non-thermal processes, such as synchrotron radiation from a relativistic jet or the emergence of emission from a young pulsar wind nebula (see the ionization shock breakout scenario in \citealt{metzger14}). Because these models predict different X-ray luminosities, temporal evolution, and spectral energy distributions, the X-ray monitoring of SLSNe can help constrain the energy source powering their luminous displays. 

With the exception of SCP\,06F6 ($z = 1.189$, \citealt{Levan13}), X-ray observations of nearby SLSNe-I, have resulted in non-detections,  with limits as deep as $L_{\rm x} \sim 10^{40} \, \rm erg\,s^{-1}$ \citep{margutti18}. X-ray emission at the location of SCP\,06F6 was reported at the level $L_{\rm x} \sim 10^{45} \, \rm erg\,s^{-1}$ (0.2 -- 2\,keV) about $\sim 70$ days (rest-frame) after the explosion date \citep{Levan13}. At this epoch, SCP\,06F6 is significantly more luminous than a Gamma-Ray Burst (GRB) SN (see \citealt{margutti18}, their Figure 7).\footnote{Another SLSN-I of note is PTF\,12dam ($z = 0.107$). In this case, an X-ray source is detected at the location of the transient with $L_{\rm x} \sim 2 \times 10^{40} \, \rm erg\,s^{-1}$ (0.3 -- 10\,keV). However, the X-ray emission is consistent with star formation from the host galaxy, there is no evidence for temporal variability, and it should be thus treated as an upper limit on the transient X-ray output \citep{margutti18}.}

In this work, we investigate SN 2018bsz, one of the nearest SLSNe-I to date ($z = 0.0267$, $d_{\rm L} = 111 \,$ Mpc; \citealt{Anderson18}). Due to its proximity, SN\,2018bsz received extensive multi-wavelength followup. The initial optical study presented in \cite{Anderson18} discusses the light curve evolution of SN\,2018bsz, highlighting a long ($> 26$\,d), slowly rising ``plateau'' before a steeper rise to the optical peak. This feature is unique among SLSNe. In the context of ordinary SNe, this long phase of emission ahead of the optical peak has been interpreted as an observational signature of outflows or other stellar activity prior to stellar explosion (e.g., \citealt{strotjohann21, wynn22}). Continued spectroscopic analysis of SN\,2018bsz identified the emergence of hydrogen emission lines between 10 to 40 days post optical peak \citep{purs22}, indicating possible interaction between the initially H-poor SN ejecta and H-rich CSM. The authors report the presence of multi-component H$\alpha$ emission beginning at $\sim 30$ days post optical peak.
In addition, the late-time infrared photometry presented by \citet{chen21} identifies significant dust formation in the ejecta of SN\,2018bsz, a phenomenon typically assumed to occur in regions of strong CSM interaction. Together, these observations provide independent indications of CSM interaction contributing significantly to the emission from SN\,2018bsz. Finally, in a recent study \citet{sankar26} present evidence for NIR-optical-UV quasi-periodic modulations on a timescale of $\sim 32$\,d in the light curve of SN\,2018bsz using observations spanning 2--110\,days post optical peak. This study suggests Lense-Thirring precession of the kind proposed for SN\,2024afav \citep{farah24afav}, interaction with a pre-existing structured CSM, or post-SN interaction between a surviving companion and a compact remnant, similar to the system proposed for SN\,2022jli \citep{chenp24}, as possible physical origins of the observed quasi-periodicity.

In this paper, we develop a simplified framework to model the thermal X-ray emission from the forward and reverse shocks of SNe, accounting for shock dynamics and photoelectric absorption in both the CSM and SN ejecta \S\ref{Sec:Model}. This framework is applied to SN\,2018bsz in \S\ref{Sec:Application}. The X-ray detection of SN\,2018bsz with our CXO campaign has been presented in \citet{ahlvind26}, who estimate the progenitor mass-loss rate from an analytic luminosity scaling for the reverse shock. We reanalyze our CXO data in \S\ref{Subsec:CXO} and present evidence for time variability in \S\ref{Sec:TimeVariability}. We place SN\,2018bsz in the context of other SLSNe-I and SN\,2022jli in \S\ref{Sec:Comparison}, and summarize our conclusions in \S\ref{Sec:Conclusions}. We note that only the CSM interaction interpretation for SN\,2018bsz is considered here and other scenarios, such as the ionization breakout, have been explored in \cite{ahlvind26}. These arguments are not repeated here.

Throughout this paper, we follow \cite{Anderson18} and assume an explosion date of March 25, 2018 (which is MJD = 58202.5). Times are referred to this date unless otherwise noted.


\begin{figure}
\centering
 \includegraphics[width=\columnwidth]{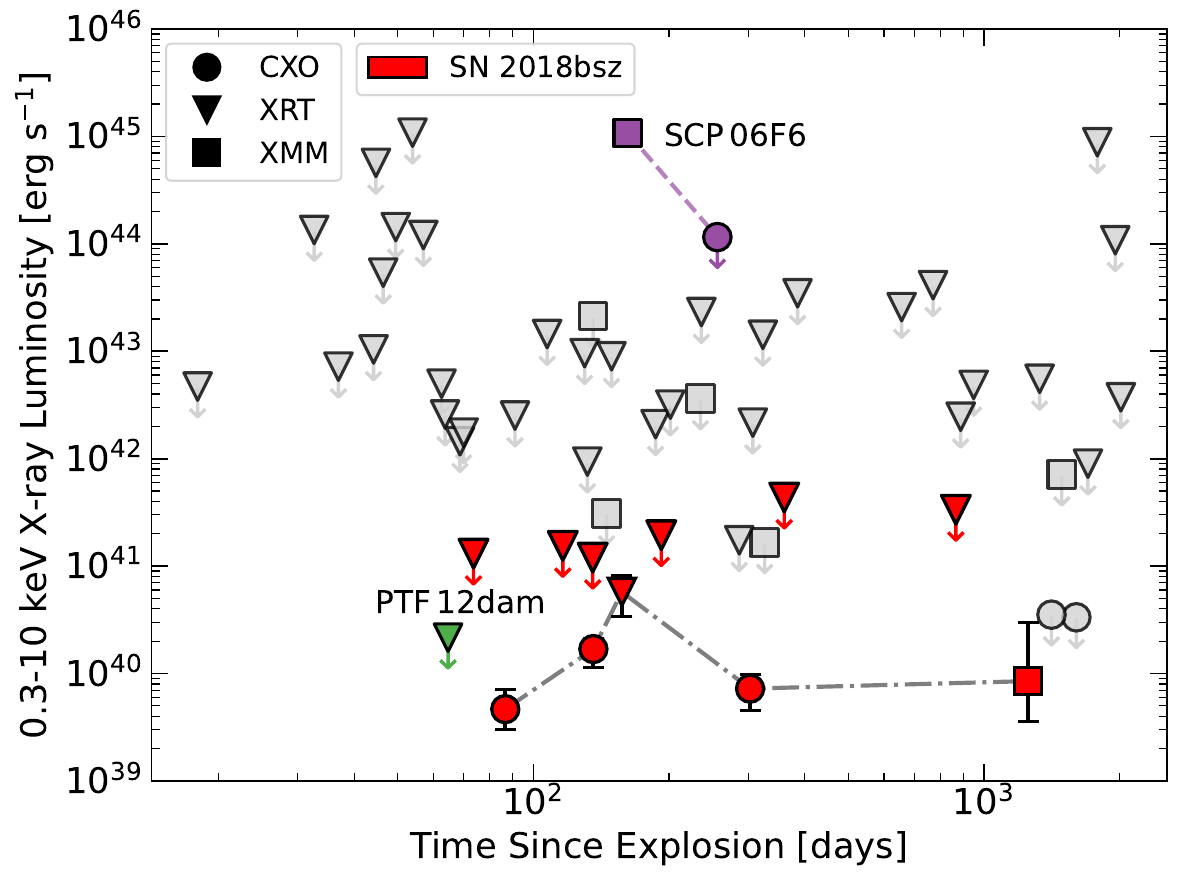}
 \caption{0.3--10\,keV X-ray luminosity evolution of SN\,2018bsz (red points) in the context of SLSNe-I from \cite{margutti18} and \cite{Levan13}. Downward arrows indicate the point is a 3$\sigma$ upper limit. The late-time XMM detection of SN\,2018bsz has been presented in \citet{ahlvind26} (here adapted to the 0.3 -- 10\,keV energy range). Notable SLSNe-I PTF\,12dam and SCP\,06F6 are in color.}
\label{Fig:slsnexrays}
\end{figure}
\noindent

\section{Observations}
\label{Sec:Obs18bsz}

\subsection{Swift-XRT (0.3--10\,keV)}
\label{Subsec:XRT}
The X-Ray Telescope (XRT; \citealt{Burrows05}) on board the Neil Gehrels Swift Observatory \citep{Gehrels04} observed SN\,2018bsz for a total of 76.9\,ks in the time period 2018-06-18 until  2020-08-07 ($\delta t =$ 85.1--866.3\,d). We process the data using \texttt{HEASoft v6.36.0} and the latest \texttt{CALDB} release following standard criteria (e.g., \citealt{Margutti13}). We find tentative evidence for a weak X-ray source at the location of the SN in ObsID 00010706022 at ($\delta t=$ 156.7\,d). No X-ray emission is otherwise detected during the XRT monitoring campaign. Although this is a targeted detection with a limited confidence level of $3.2\sigma$ (Gaussian equivalent), it is interesting to note that the XRT detection corresponds to the time of peak of the X-ray light curve obtained with the Chandra X-ray Observatory. We thus consider possible that the XRT detection is not a mere statistical fluctuation and it is instead related to the transient. For the flux calibration, we assume a spectrum with photon index in the range $\Gamma=1-1.5$ with local and Galactic absorption as derived in \S\ref{Subsec:CXO}. From these parameters we infer $L_x=(5.7\pm 2.3)\times 10^{40}\,\rm{erg\,s^{-1}}$ (0.3--10\,keV).

\subsection{CXO (0.3 -- 10\,keV)}
\label{Subsec:CXO}
We observed SN\,2018bsz at three different epochs with the Chandra X-ray Observatory (CXO; PI: Margutti). The first epoch was acquired on 2018-06-20 ($\delta t \approx 87$\,d rest-frame since explosion, $49.6$\,ks exposure time), followed by a second epoch split into two observations on 2018-08-01 and 2018-08-14 ($\delta t \approx 129$ and $\delta t \approx 142$\,d rest-frame since explosion, $25.9$ and $20.7$\,ks exposures, respectively). The third epoch was observed on 2019-01-22 ($\delta t \approx 303$\,d rest-frame since explosion, $49.4$\,ks exposure). The data were reprocessed using the software package {\tt CIAO} (v4.17), and standard ACIS data filtering was applied.

An X-ray source is blindly detected by {\tt wavdetect} at a location statistically consistent with the optical position of SN\,2018bsz in the 0.5--8\,keV filtered data for each epoch. The statistical significance of each detection and the associated net count-rates are reported in Table \ref{Tab:18bsz_xrays}. We note that this is a faint source for which Poisson statistics applies.

For each observation, we use {\tt specextract} to extract an X-ray spectrum from a $2\arcsec$ source region located at each wavdetect centroid and a $35\arcsec$ source-free background region. The Galactic neutral hydrogen column density in the direction of SN\,2018bsz is $N_{\rm MW}=1.22\times 10^{21}\,\rm{cm^{-2}}$ \citep{HI4PI}. To estimate the intrinsic neutral hydrogen column density ($N_{\rm int}$), we begin by considering the optical extinction towards SN\,2018bsz. We adopt the value $E(B-V)_{\rm host}=0.04$\,mag inferred by \cite{Anderson18} from the equivalent width of the Na D absorption lines within the host galaxy \citep{Poznanski12}. This corresponds to $A_{V, \rm \,host} = 0.13$\,mag, assuming $R_{V, \rm \, host} = 3.1$ (\citealp{Cardelli89}). Assuming a Galactic dust-to-gas ratio, we obtain $N_{\rm int} \approx 2.3\times 10^{20}\,\rm{cm^{-2}}$ using the $N_{\rm int}/A_V$ correlation from \cite{Predehl95} and $N_{\rm int} \approx 2.9\times 10^{20}\,\rm{cm^{-2}}$ using the correlation from \cite{Watson11}. The estimates above point to minimal intrinsic absorption. In the following, we adopt the central value $N_{\rm int}=2.6\times 10^{20}\,\rm{cm^{-2}}$ as a fiducial estimate, and we note that our main results do not depend on this choice.

Using these values, we model the spectrum for each epoch using the software package {\tt Xspec} (v12.14.1) with an absorbed power-law model. We performed a joint spectral fit for the split data sets of the second epoch, since we do not expect significant source variability on a timescale of $\sim13$\,d. We estimate $1\sigma$ errors in {\tt Xspec} using a Markov chain Monte Carlo fit calculated with the Goodman-Weare algorithm. We initialize 10 walkers over a chain length of $10^4$, and report our results in Table \ref{Tab:18bsz_xrays}. We find that our obtained flux values are in statistical agreement with those presented in the recent study by \citet{ahlvind26}. A comparison between the 0.3--10\,keV band X-ray luminosities outlined in this section and \S\ref{Subsec:XRT} against other SLSNe-I upper limits is shown in Figure \ref{Fig:slsnexrays}.

\section{Modeling and Discussion}
\label{Sec:Analysis}
We begin this section with a deeper investigation into the origin of the X-ray emission detected at the location of SN\,2018bsz. We then develop a model for the X-ray emission from the SN shocks interaction. Finally, we apply the shock interaction model to SN\,2018bsz. We end with a comparison to other X-ray SNe. 

\bigskip
\subsection{Time Variability}
\label{Sec:TimeVariability}
We use a binomial test to evaluate whether the observed variability in X-ray emission detected by Chandra in the direction of SN\,2018bsz represents a random fluctuation of an intrinsically constant source. For this analysis, we make separate comparisons between Epochs 1 and 2, and Epochs 2 and 3. We take the total number of photon counts in each epoch to be the sum of the source and background counts in a $2\arcsec$ region in the 0.5--8\,keV range (see Table \ref{Tab:18bsz_xrays}). We calculate a probability of $\sim 0.3\%$ ($\approx3\,\sigma$) for brightening by chance between Epochs 1 and 2, and a probability of $\sim 2\%$ ($\approx2.3\,\sigma$) for fading by chance between Epochs 2 and 3. We thus conclude that there is tentative evidence for intrinsic variability of the source at the $3\,\sigma$ level. Additionally, due to the correspondence between the XRT detection and the time of peak in the Chandra light curve, we consider it plausible that this emission is related to the transient. These results suggest that the X-ray emission in the direction of SN\,2018bsz is not constant, but instead variable over the timescale of a few weeks.\footnote{Note that \cite{ahlvind26} concluded instead that the source is consistent with being constant as the analysis was performed in the flux space-phase and did not fully account for the Poissonian nature of the signal.}

\subsection{Star Formation}
\label{Sec:StarFormation}
There is an expected correlation between the star formation rate (SFR) of a galaxy and its associated X-ray emission due to X-ray binaries (XRBs; dominant in the 2--10\,keV energy range), and the hot interstellar medium (ISM; dominant in the 0.5--2\,keV energy range; see e.g., \citealt{Ranalli03,Kennicutt12,Mineo12} for a review). We use these correlations to quantify the possible contribution of star formation to the X-rays detected at the location of SN\,2018bsz. We adopt the host-galaxy SFR derived by \cite{Anderson18} for this analysis, $\rm{SFR} = 0.50 \substack{+2.22 \\ -0.19} \,M_\odot \,\rm{yr}^{-1}$.

From \S\ref{Subsec:CXO}, the X-ray luminosity at the location of the source over Epoch 1 is $L_{\rm{x}} \approx 1.2\times 10^{39}\,\rm{erg\,s^{-1}}$ (0.5--2\,keV). The relations between SFR and X-ray emission due to hot ISM derived in \cite{Mineo12} (their Equations 1 and 2: $L_{\rm{x}}\,/\,\rm{SFR} \approx 8.3\times 10^{38}\,\rm{erg\,s^{-1}}\,(M_\odot\,\rm{yr}^{-1})^{-1}$ and $L_{\rm{x}}\,/\,\rm{SFR} \approx 5.2\times 10^{38}\,\rm{erg\,s^{-1}}\,(M_\odot\,\rm{yr}^{-1})^{-1}$, respectively) predict $L_{\rm{x}} \approx (0.3 - 0.4)\times 10^{39}\,\rm{erg\,s^{-1}}$ (0.5--2\,keV). This result is roughly $\approx 1/3$ of our measured value.

Similarly, we determine the X-ray luminosity of SN\,2018bsz in Epoch 1 to be $L_{\rm{x}} \approx 3.3\times 10^{39}\,\rm{erg\,s^{-1}}$ (2--10\,keV). We use the relationship $\log\,L_{\rm{x}} = A + B\,\log\,{\rm{SFR}} + C\,\log(1+z)$ presented in Equation 14 of \citet{Lehmer16}, where $A$, $B$, and $C$ are fitting constants\footnote{We take $A = 39.82$, $B = 0.63$, and $C = 1.31$ in the 2 -- 10\,keV band.} in the 2--10\,keV band. This predicts $L_{\rm{x}} \approx 3\times 10^{39}\,\rm{erg\,s^{-1}}$ (2--10\,keV), which is similar to our measurements.

From these results, we conclude it to be possible that star formation in the host galaxy of SN\,2018bsz contributes to at least some of the detected X-ray emission at the location of the transient. We note that estimates of the X-ray luminosity in additional energy bands other than those reported here lead to the same conclusion. However, from \S\ref{Sec:TimeVariability}, there is tentative evidence that this source is variable on timescales that are much shorter than expected in the star formation scenario. Together with the large uncertainties associated with the calibration of the SFR--$L_{\rm{x}}$ relationships and the evidence of SN shock interaction from optical spectroscopy and photometry (see e.g., \citealt{Anderson18, chen21}), which is timed with the detection of the X-ray emission, we thus consider plausible that the detected X-rays are dominated by transient emission.

\subsection{CSM Interaction Modeling}
\label{Sec:Model}
In this section, we present the main contribution of this work, a semi-analytic model for the X-ray radiation produced as a result of the deceleration of the SN shocks in the environment.
Broadly speaking, we follow the approach of \cite{wiston26} with regard to the shock dynamics, and \cite{ibik25} for the radiation. We note, however, that we are unable to reproduce the results shown in Figure 16 of \cite{ibik25} due to their inaccurate treatment of the CSM absorption that we correct and detail in Appendix \ref{Sec:appendixA}.

\subsubsection{Shock Dynamics}
\label{SubSubSec:dynamics}
First, we consider the interaction of the SN ejecta with a pre-existing CSM created by progenitor mass-loss via stellar winds. In this wind model, the CSM is taken to be stationary, spherically symmetric, and without clumps or asymmetries. For a wind-like scenario, the density of the CSM is described with a power-law relationship following the scaling $\rho_{\rm CSM}\propto R_{\rm sh}^{-s}$, where $s=2$. Specifically, we have:

\begin{equation}
    \rho_{\rm CSM} = \frac{\dot{M}}{4\pi v_{\rm w} R_{\rm sh}^2}
    \label{eq:rho_csm}
\end{equation}
where $\dot{M}$ is the mass-loss rate of the progenitor star, $v_{\rm w}$ is the wind velocity, and $R_{\rm sh}$ is the shock radius. Both $\dot{M}$ and $v_{\rm w}$ are taken to be constant over time, corresponding to steady progenitor mass loss and a constant wind velocity.

The ejecta density structure derived in \cite{chev89} for homologously expanding SN ejecta is described by a broken power-law profile consisting of a shallow inner region described by $\delta$ and a steep outer region described by $n$ in the velocity $v$ space:

\begin{equation}
\rho_{\rm ej}(v) = 
    \begin{cases} 
        F\left(v/v_{\rm t}\right)^{-\delta}\, t^{-3}, & v < v_{\rm t} \\[1em]
        F\left(v/v_{\rm t}\right)^{-n} \, t^{-3}, & v \geq v_{\rm t}
    \end{cases}
    \label{eq:rho_ej}
\end{equation}

where
\begin{equation}
    F \equiv \frac{1}{4\pi(n-\delta)} \frac{[(3-\delta)(n-3)M_{\rm ej}]^{5/2}}{[2(5-\delta)(n-5)E_{\rm k}]^{3/2}},
\end{equation}

\begin{equation}
    v_{\rm t} \equiv \left[\frac{2(5-\delta)(n-5)}{(3-\delta)(n-3)} \frac{E_{\rm k}}{M_{\rm ej}} \right]^{1/2}.
    \label{eq:v_transition}
\end{equation}
Here, $v_{\rm t}$ is the transition velocity separating the inner and outer power-law regimes, and $M_{\rm ej}$, $E_{\rm k}$ are the total mass and total kinetic energy of the initial explosion, respectively.

We begin by solving for the shock dynamics in the thin shell approximation (e.g., \citealt{chev82}). In this framework, the interaction region is approximated as a geometrically thin shell characterized by a single shock radius $R_{\rm sh}$ and shock velocity $v_{\rm sh}$. The conservation of momentum equation for the shocked shell of CSM and ejecta material is written as:
\begin{equation}
    M_{\rm sh} \frac{dv_{\rm sh}}{dt} = 4\pi R_{\rm sh}^2[\rho_{\rm ej}(v_{\rm ej} - v_{\rm sh})^2 - \rho_{\rm CSM}v_{\rm sh}^2]
    \label{eq:shockevo}
\end{equation}
where $M_{\rm sh}$ is the total mass of the shocked ejecta and shocked CSM material, $v_{\rm sh} = dR_{\rm sh}/dt$ is the shock velocity, $v_{\rm ej} = R_{\rm sh}/t$ is the velocity of the ejecta crossing the shock front, and $\rho_{\rm ej}$, $\rho_{\rm CSM}$ are the density profiles of the ejecta and CSM, respectively (e.g., \citealt{chevalier17}). We numerically integrate Equation \ref{eq:shockevo} using a fourth-order Runge–Kutta method with $n$, $\delta$, $E_{\rm k}$, $M_{\rm ej}$, and $v_{\rm w}$ as inputs. Numerical integration requires the use of initial values $v_{\rm 0}$ and $R_{\rm 0}$ at a specified starting time $t_{\rm 0}$. Under the adopted assumptions, the resulting solution rapidly approaches self-similar evolution and is thus insensitive to these arbitrary starting values.

\subsubsection{X-ray Emission from CSM Shock Interaction}
\label{SubSubSec:radiation}
The conversion of kinetic energy into thermal energy at the interaction region generates X-ray emission via thermal bremsstrahlung. We expect bremsstrahlung emission to dominate the X-ray spectrum due to the high densities involved in the CSM interaction model \citep{chev98,Chevalier06}.

From \cite{Fransson96}, the parameters of the shocked material in the strong shock limit ($\gamma = 5/3$) are given as follows:
\begin{subequations}
\begin{align}
    \rho_{\rm FS} & = 4\rho_{\rm CSM} &
    M_{\rm FS} & = \frac{\dot{M}R_{\rm sh}}{v_{\rm w}} \label{eqn:fsprop}\\
    \rho_{\rm RS} & = \frac{(n-4)(n-3)}{(3-s)(4-s)} \cdot \rho_{\rm FS} &
    M_{\rm RS} & = \frac{(n-4)}{(4-s)} \cdot M_{\rm FS} \label{eqn:rsprop}
\end{align}
\end{subequations}
where the expressions in Equation (\ref{eqn:fsprop}) are the density and mass of the shocked CSM material, and the expressions in Equation (\ref{eqn:rsprop}) are the density and mass of the shocked ejecta material, respectively. ``FS'' (``RS'') stands for Forward (Reverse) Shock. 

The conservation of energy, mass, and momentum across the shock front  implies an ion temperature of (see e.g., \citealt{chevalier17}):
\begin{equation}
    T_\lambda = 2 \frac{(\gamma-1)}{(\gamma+1)^2} \frac{\mu_\lambda m_{\rm H}}{k_b} v_\lambda^2
    \label{eq:temp}
\end{equation}
where $\mu_{\lambda}$ is the mean mass per particle in the shocked region of interest, $v_\lambda$ is the shock velocity (where $\lambda$ denotes either FS or RS), and $m_{\rm H}$ is the proton mass. From our definitions it follows that $v_{\rm FS} = v_{\rm sh}$ and $v_{\rm RS} = R_{\rm sh}/t - v_{\rm sh}$.
In the following, we assume electron-ion equilibrium (i.e., $T\equiv T_e$), and we check for self-consistency of our treatment for SN\,2018bsz in \S\ref{Sec:Application}. 

The unabsorbed free-free luminosity of each shock is estimated under the assumption of spherical symmetry as \citep{Fransson96}:
\begin{equation}
    L_{\lambda}(\nu) = 4\pi\int j_{\rm ff}(\nu)\Big (\frac{\rho}{m_{\rm H}}\Big )^2
    r^2 dr \approx j_{\rm ff}(\nu) M_{\lambda} \, \rho_{\lambda} /m_{\rm H}^2,
    \label{eqn:xlum}
\end{equation}
where $M_{\lambda}$, $\rho_{\lambda}$ are the mass swept up by the shock and the density of the shocked material, respectively. We note that the last approximation is only valid in the limit of an adiabatic shock (while for a radiative shock the emitting region should be reduced to the volume of mass that crossed the shock during a cooling time). This approximation is adequate for our purposes of modeling SN\,2018bsz (see \S\ref{Sec:Application}).
The emissivity, $j_{\rm ff}(\nu)$, is a function of spectral frequency, and depends on the temperature of the shocked electrons ($T_{\rm e}$) as:
\begin{equation}
    j_{\rm ff}(\nu) = 1.64 \times 10^{-20} \zeta g_{\rm ff} T_{\rm e}^{-0.5} e^{-h\nu/k_b T_{\rm e}}~\rm{erg\, cm^3\,s^{-1}\,keV^{-1}},
\end{equation}
where $\zeta$ is a composition factor and $g_{\rm ff}$ is the free–free Gaunt factor. From \cite{Fransson96}, we use $\zeta = \big[1+2\cdot n({\rm He})/n({\rm H})\big] / \big[1+4\cdot n({\rm He})/n({\rm H}) \big]$, where $n(\rm{He})$ and $n(\rm{H})$ are the densities of He and H atoms, respectively, and $g_{\rm ff} = 1.87(T_\lambda/10^8 \, \rm K)^{0.264}$. We note that this solution for $g_{\rm ff}$ holds for energies $h\nu\approx$1\,keV and is an acceptable approximation for our modeling purposes. We integrate the specific luminosity of Equation \ref{eqn:xlum} across the energy band of interest (0.3--10\,keV) to directly compare to  luminosities inferred from observations. Additionally, we account for the fact that only $\approx$ half of the luminosity of Equation \ref{eqn:xlum} reaches the observer due to ejecta occultation at these early times \citep{Fransson96, margalit18}.

\begin{figure*}
\centering
 \includegraphics[width=\textwidth]{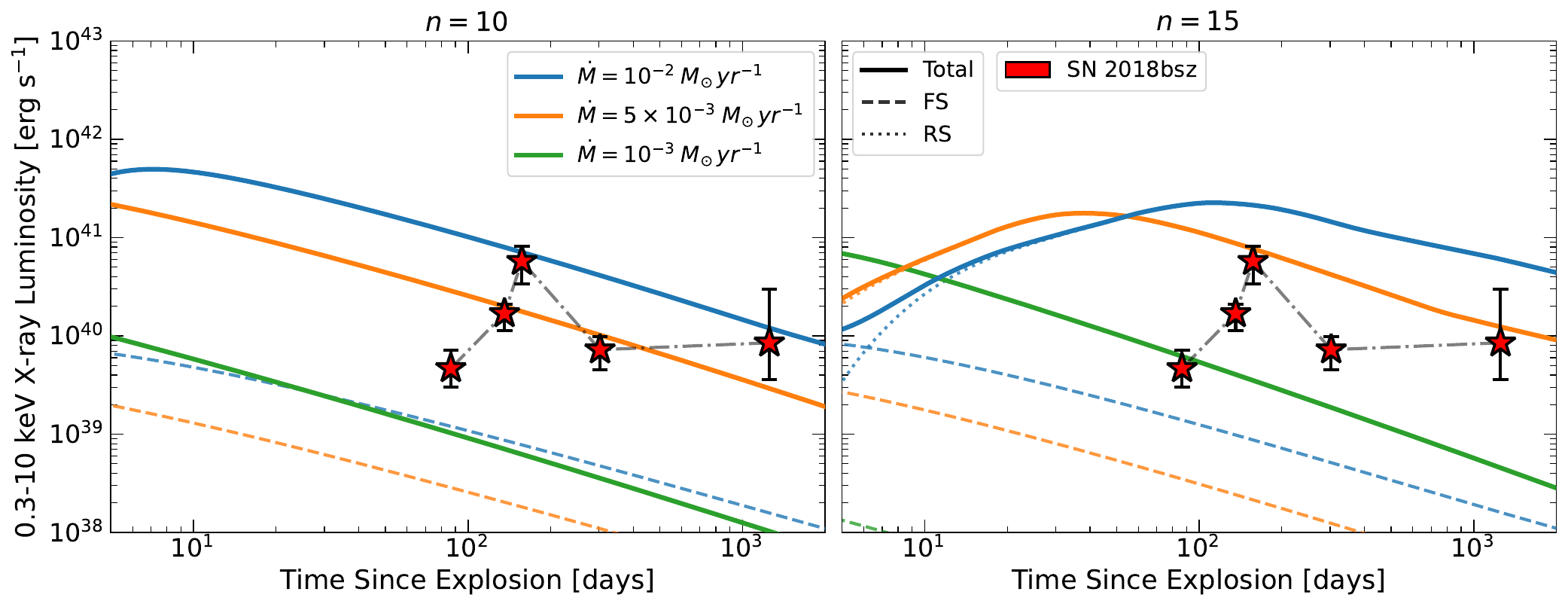}
 \caption{Predicted evolution of the absorbed 0.3--10\,keV X-ray luminosity for the FS (dashed lines) and RS (dotted lines) for a range of $\dot M$ values using the input parameters $E_{\rm k} = 10^{52}$ erg and $M_{\rm ej} = 10 \, \rm M_\odot$ similar to those derived by \citet{gomez2024} from the modeling of the optical light curve. Solid line: total (FS+RS) absorbed X-ray emission. The steeper progenitor outer density profile $n=15$ (right panel) leads to an earlier peak in the X-ray emission than in the progenitor case with $n=10$ (left panel). In both cases, however, we find the higher mass-loss cases to be broadly consistent with the level of X-ray emission detected for SN\,2018bsz (red stars).}
\label{Fig:10msunmodeling}
\end{figure*}
\noindent

\begin{figure*}  
\centering
 \includegraphics[width=\textwidth]{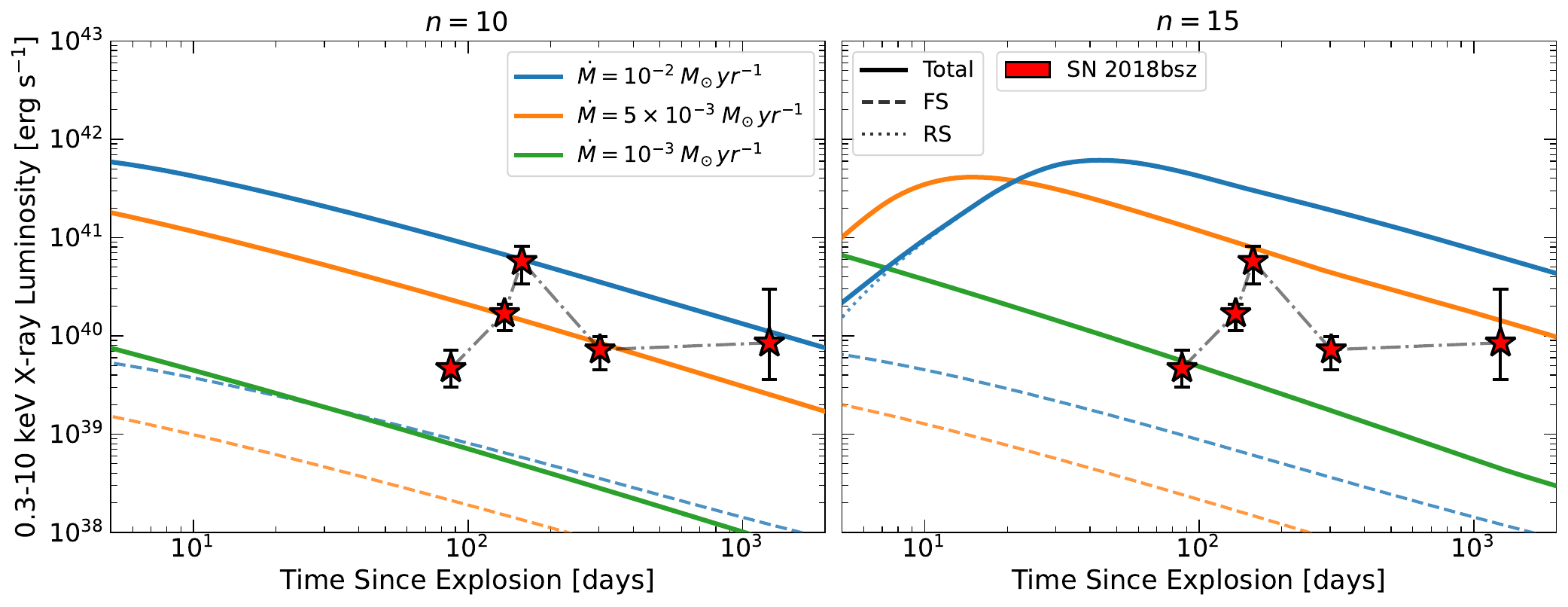}
 \caption{Predicted evolution of the absorbed 0.3--10\,keV X-ray luminosity using the input parameters $E_{\rm k} = 10^{52}$ erg and $M_{\rm ej} = 5 \, \rm M_\odot$ similar to those derived by \citet{Anderson18}. Same color coding as Figure \ref{Fig:10msunmodeling}.}
\label{Fig:5msunmodeling}
\end{figure*}
\noindent

Material in front of the shocks can lead to photoelectric absorption of the X-ray emission, with cold, neutral material making for a more efficient absorber than a hot, ionized medium. The shocked material can cool either radiatively through bremsstrahlung emission or line emission, or adiabatically, due to expansion. The cooling timescale for bremsstrahlung continuum emission is (see e.g., \citealt{nymark06}, their Equation 37):
\begin{equation}
    t_{\rm cool,\lambda} = \frac{3k_b T_\lambda}{2(1-\mu_{\lambda}/\mu_{\rm {A},\lambda})\cdot n_{\rm{i},\lambda} \Lambda(t)}
    \label{eqn:tcool}
\end{equation}
where $\mu_{\rm{A},\lambda}$ is the mean \emph{atomic} weight and $n_{\rm{i},\lambda}$ is the number density of ions. The cooling function, including cooling via bremsstrahlung and line emission, is parametrized by \cite{nymark06} as $\Lambda(t) = AT_6^{-\alpha} + BT_6^\beta$ with $T_6 \equiv (T_\lambda/10^6 \, \rm K)$. Here, $A$, $B$, $\alpha$, and $\beta$ are fitting parameters dependent on the chemical composition of the gas. When $t_{\rm cool}\gg t$, the shocked gas primarily evolves adiabatically; and when $t_{\rm cool}\ll t$, radiative cooling via continuum (bremsstrahlung) and/or line emission becomes important. We note that continuum cooling dominates over line cooling for shock temperatures above $10^7$\,K. 

Cooling can enhance the fraction of neutral material, and hence the X-ray photoelectric absorption by metals. Contributions to the X-ray absorption include the unshocked CSM material upstream of the forward shock, impacting emission from both the forward and reverse shocks; and the formation of a Cold Dense Shell (CDS) in between the shocks, impacting emission from the RS. A cold dense shell is expected to form when the RS is radiative and absent otherwise. 

We estimate the column density of the unshocked CSM by integrating the CSM density profile from $R_{\rm sh}$ out to a maximum radius of $\approx10^{18} \,\rm{cm}$:
\begin{equation}
    N_{\rm CSM} = \frac{1}{\mu_{\rm A,FS} m_{\rm H}}\int^{R_{\rm out}}_{R_{\rm sh}}{\rho_{\rm{CSM}}}(r)\,dr.
    \label{eqn:ncsm}
\end{equation}
We note that the value of $R_{\rm out}=10^{18} \,\rm{cm}$ is an arbitrary choice for the purposes of numerical integration, and choosing any larger number does not have an effect on the results of our analysis due to the smaller density contributions at larger radii.\footnote{We note that treating the effects of a time-variable ionization of the CSM in front of the FS is beyond the scope of this paper and will be included in a future version of the code.}

A CDS can form as a result of efficient cooling and compression of material between the shocked ejecta and the contact discontinuity. In this case, the total absorbing column density impacting emission from the RS is $N_{\rm CSM}+N_{\rm CDS}$. Under the assumption of uniformly distributed shocked material, we define
\begin{equation}
    N_{\rm CDS} = \frac{M_{\rm RS}}{4\pi R_{\rm sh}^2\mu_{\rm A,RS} m_{\rm H}}
    \label{eqn:ncds}
\end{equation}
following Equation 50 from \cite{nymark06}. 

The absorbed X-ray luminosity is calculated independently for each shock as $L_{\rm x,\, abs} = e^{-N_{\rm H} \sigma}L_{\rm x}$, where $N_{\rm H}$ is the effective neutral hydrogen absorption column given by a combination of Equations \ref{eqn:ncsm} and \ref{eqn:ncds} above, in addition to both the Galactic and host-intrinsic absorption columns derived in \S\ref{Subsec:CXO}. Additionally, $\sigma(E)$ (where $E=h\nu$) is the photoelectric absorption cross-section as a function of energy. Since the photoelectric cross-section depends on the chemical composition of absorbing material, we consider two limiting cases: $\sigma_{\rm ISM}$ and $\sigma_{\rm ej}$. The first case, $\sigma_{\rm ISM}$, assumes an ISM-like composition defined by the abundances detailed in \cite{wilm00}. This case applies to $N_{\rm CSM}$, $N_{\rm int}$, and $N_{\rm MW}$. The second case, $\sigma_{\rm ej}$, assumes a typical SN ejecta-like composition with $\sim35\%$ O and $\sim2\%$ Fe per mass (e.g., \citealt{chugai06}), which is relevant to $N_{\rm CDS}$.

To obtain a smooth transition between the radiative and adiabatic cooling regimes of the RS, we apply the following weighting scheme, used by \cite{eftek18} in another context, to calculate the effective column density affecting the RS emission:
\begin{equation}
    N_{\rm RS} = \frac{w_1N_{\rm rad} + w_2N_{\rm ad}}{w_1 + w_2}.
\end{equation}
Here, $N_{\rm rad} = N_{\rm CSM}+N_{\rm CDS}$ (in the radiative case), $N_{\rm ad} = N_{\rm CSM}$ (in the adiabatic case), $w_1 = (t/t_{\rm cool, RS})^2$, and $w_2 = (t_{\rm cool, RS}/t)^2$.\footnote{A recent study presented in \citet{shunke26} addresses a similar transition between adiabatic and radiative regimes in the shock. Following Equation 30 in this paper, the authors consider a transition from a prescription of a single-temperature bremsstrahlung spectrum (adiabatic case) to one of a cooling bremsstrahlung spectrum (radiative case). This transition occurs via a ``radiation efficiency" term dependent on the dynamical, bremsstrahlung cooling and Compton cooling timescales of the shocked material. We note this approach to be another possible method to model the transition between regimes.}

\subsection{Application to SN\,2018bsz}
\label{Sec:Application}
The CSM interaction model predicts a thermal bremsstrahlung spectrum. For $k_bT_{\rm e}>10$\,keV that are relevant to our problem, this spectrum manifests as a hard power-law in the 0.3--10\,keV band accessible to soft X-ray spacecraft, which is consistent with our inferred spectral index of Table \ref{Tab:18bsz_xrays} given the limited CXO statistics. We thus consider the CSM interaction model viable from a ``spectral shape'' perspective and proceed further with our modeling.
We consider steady progenitor mass-loss rates in the range $\dot{M} = 10^{-6}$ to $\dot{M} = 10^{-2}\, \rm M_\odot\, yr^{-1}$, for an assumed constant wind velocity of $v_{\rm w} = 1000 \,\rm km\,s^{-1}$, appropriate for Wolf--Rayet (WR) progenitors. Since our modeling is sensitive to the CSM \emph{density}, $\rho_{\rm{CSM}}\propto \dot M/v_{\rm w}$, our inferred $\dot M$ can be easily rescaled for any other choice of $v_{\rm w}$. Additionally, in the following, our model assumes spherical symmetry of the interaction region. However, \cite{purs22} inferred an asymmetric CSM for SN\,2018bsz. We note that the inferred $\dot M$ below scale as $\propto 1/f^{0.5}$, where $f<1$ is the volume filling factor of the emitting region.

The density of the ejecta material is parametrized using the broken power-law profile described in Equation \ref{eq:rho_ej} with inner slope $\delta = 2$. We consider the cases $n = 10$ and $n = 15$ in order to explore a range of outer density profiles. The $n \approx 10$ case represents the ejecta structure expected for a compact WR-like progenitor, and $n \approx 15$ represents the ejecta structure expected for an extended supergiant-like progenitor. For the CSM gas composition, we assume solar abundances and adopt $\mu_{\rm FS} = 0.61$, $\mu_{\rm A,FS} = 1.29$, $A = 8.0 \times 10^{-23}\, \rm{erg\, cm^3\, s^{-1}}$, $B = 2.3 \times 10^{-24} \, \rm{erg\, cm^3\, s^{-1}}$, $\alpha = 0.90$, and $\beta = 0.50$ from Tables 1 and 2 of \cite{nymark06}. Using $n({\rm He})/n({\rm H})=0.1$ for a solar-composition gas \citep{Fransson96}, we take $\zeta = 0.86$. For an ejecta-like gas composition, we take $\mu_{\rm RS} = 1.44$ and $\mu_{\rm A, RS} = 5.41$. 

We model the evolution of the absorbed X-ray luminosity over time for two choices of total kinetic energy ($E_{\rm k}$) and mass of the SN ejecta ($M_{\rm ej}$). We note that the shock dynamics are primarily sensitive to the ratio between these properties, which determines the initial shock speed. From the modeling of the optical light curve of SN\,2018bsz with \texttt{MOSFiT} \citet{gomez2024} infer $E_{\rm k} = 1.2^{+1.1}_{-0.6}\times10^{52}$ erg and $M_{\rm ej} = 9.5^{+7.5}_{-4.4} \, \rm M_\odot$. \cite{Anderson18} derive similar $E_{\rm k}$ values with a slightly reduced mass $M_{\rm ej} = 5.6\, \rm M_\odot$. We thus adopt two sets of values to represent the uncertainty associated with these estimates, and we show the resulting 0.3--10\,keV absorbed X-ray light curves in Figure \ref{Fig:10msunmodeling} ($E_{\rm k} = 10^{52}$ erg, $M_{\rm ej} = 10 \, \rm M_\odot$), and Figure \ref{Fig:5msunmodeling} ($E_{\rm k} = 10^{52}$ erg, $M_{\rm ej} = 5 \, \rm M_\odot$). Our main conclusion from this exercise is that, generally speaking, we find that mass-loss rates in the range $\dot M\approx 10^{-3}-10^{-2}\,\rm{M_{\odot}yr^{-1}}$ can adequately explain the X-ray luminosities inferred at the location of SN\,2018bsz.

A few considerations follow. First, we check for self-consistency of our treatment. To self-consistently check for equipartition between electrons and ions in the shocked regions, we begin by considering the equipartition timescale (Equation 12 from \citealt{chevalier17}):
\begin{equation}
    t_{\rm eq} \approx 2.5 \times 10^7\,{\rm s}~\bigg(\frac{T_{\rm e}}{10^9\,\rm{K}}\bigg)^{1.5} \bigg(\frac{n_{\rm e}}{10^7\,\rm{cm}^{-3}}\bigg)^{-1}.
    \label{eq:teq}
\end{equation}
In this equation $T_{\rm e}$ is the electron temperature and $n_{\rm e}$ is the electron number density, calculated as $n_{\rm e} = \rho_\lambda/\mu_{\rm e} m_{\rm H}$ under the assumption of a solar composition for the FS ($\mu_{\rm e,FS} = 1.14$) and ejecta composition for the RS ($\mu_{\rm e,RS} = 1.96$). We find $t_{\rm eq} \leq t$ at all times for the RS but not necessarily for the FS. At the time of interest for SN\,2018bsz the X-ray emission is dominated by the RS (e.g., Figures \ref{Fig:5msunmodeling}, \ref{Fig:10msunmodeling}), and electrons that deviate from equipartition will be cooler than our FS prediction (shifting their emission to lower energies): we thus consider our approximation adequate for our purposes. 

Additionally, using the above parameters, we check the adiabaticity of the FS and RS. Calculating $t_{\rm cool}$ using Equation \ref{eqn:tcool} reveals that $t_{\rm cool, FS} \gg t_{\rm{dyn}}$ in all cases for $\delta t > 1$\,d, showing that the FS remains adiabatic throughout the considered time frame.  We perform a similar calculation for the RS, and we show the ratio $t_{\rm{cool, RS}}/ t_{\rm{dyn}}$ in Figure \ref{Fig:tcoolRS}: we find $t_{\rm{cool, RS}}/ t_{\rm{dyn}}>1$  at all times of our monitoring. From this, we conclude that accounting for the cooling of the RS at early times in our simulations is thus justified.

Second, we comment on general trends of the modeled $L_{\rm x}$. At very early times of a few days the X-ray luminosity can be dominated by fading emission by the FS. As the absorption by the CDS in front of the RS becomes less severe with time, the X-ray emission from the RS emerges and eventually dominates the overall luminosity budget. This leads to the commonly observed ``fast fading plus re-brightening'' behavior in X-ray SNe (see, e.g., \citealt{Dwarkadas25}). However, we note that more complex $L_{\rm x}$ behaviors are possible, even without invoking deviations from spherical symmetry or complex $\rho_{\rm{CSM}}$. For example, in each of the models involving the steep ejecta density profile ($n = 15$ case), we see that the RS emission begins to diverge from monotonic evolution at late times (this is especially evident in the higher mass-loss cases). This effect is due to the ejecta velocity falling below the transition velocity, $v_{\rm t}$ in Equation \ref{eq:v_transition}, causing a change in the dynamics of the RS as it begins to propagate through ejecta material defined by the shallower power-law density core.

Third, we comment on the emergence of H$\alpha$ emission in SN\,2018bsz at $\approx 75-105$\,d post explosion \citep{purs22}. The H$\alpha$ line shows a complex profile (their Figure 4) that evolves to lose nearly all of its blue-shifted wing. This profile is distinct in shape from the ``boxy'' H$\alpha$ typically observed in strongly interacting SNe (e.g., 2023ixf, \citealt{Folatelli25,JacobsonGalan25}) for which this H emission is associated with ejecta material in the CDS. The late appearance of the H emission implies that the H-rich material in SN\,2018bsz is instead most likely CSM-related. For these reasons, we find no tension between the predominantly adiabatic reverse shock inferred here and the concurrent appearance of H$\alpha$ emission. 

To conclude: from our analysis, we find that the CSM interaction model can potentially explain the overall evolution of the $L_{\rm x}$ from SN\,2018bsz. Although our model cannot reproduce short-term X-ray variability by design, we find that a wind-like medium with mass-loss rate in the range $\dot{M} = 10^{-3}-10^{-2}\, \rm M_\odot\, yr^{-1}$ (for assumed $v_{\rm w} = 1000 \,\rm km\,s^{-1}$) can adequately reproduce the observed trend, including the lack of detectable fading of the signal at $\delta t>1000$\,d implied by the XMM detection reported by \citet{ahlvind26}. 
We note that no single steady-wind model reproduces the shape of the X-ray light curve: the rise between $\sim 87$ and $\sim 150$ days and the steep subsequent decline require $\dot M$ to vary with radius. This suggests that a non-negligible portion of the X-ray emission may arise from a localized density enhancement, consistent with the aspherical, confined CSM inferred by \citet{purs22}. Moreover, we note that X-ray variability can be plausibly produced by clumps in the CSM and/or emission from multiple outflows.
The range of inferred $\dot M$ is comparable to the estimates of \citet{ahlvind26} for the same $v_{\rm w}$, which are based on an analytic RS luminosity scaling with no attempt to do a self-consistent modeling of the dynamics and/or absorption. At the epochs of our CXO observations, the X-ray emission in our model is dominated by the reverse shock.
Continued deep X-ray monitoring of SN\,2018bsz on time scales of years can confirm or rule out this interpretation.

\subsection{Comparison to other SLSNe-I and the ``periodic'' SN\,2022jli}
\label{Sec:Comparison}

\begin{figure}[t!]
\centering
 \includegraphics[width=\columnwidth]{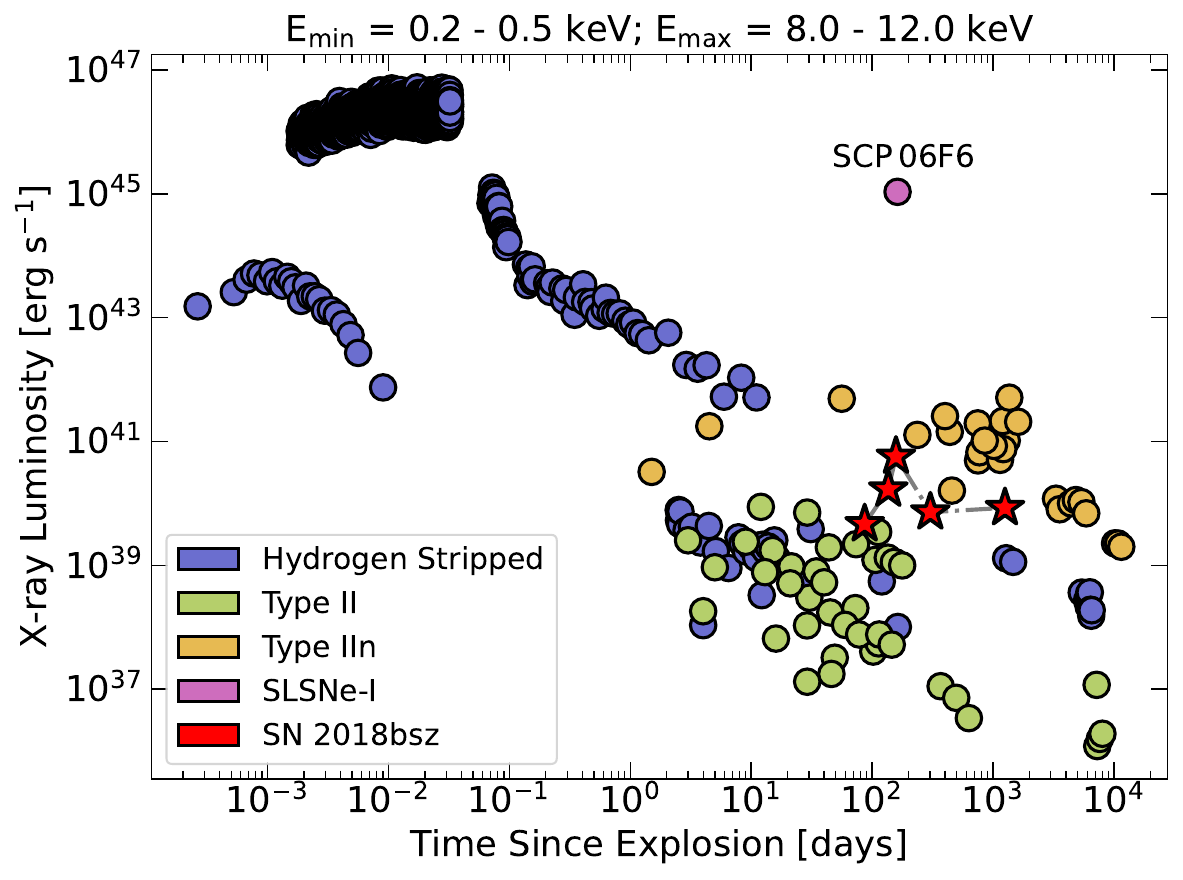}
 \caption{SN\,2018bsz (red stars) in the context of X-ray SNe of different spectral types. The data plotted here were acquired in an X-ray energy band between $E_{\rm{min}}$ and $E_{\rm{max}}$, with $E_{\rm min} =$ 0.2--0.5\,keV, and $E_{\rm max} =$ 8.0--12.0\,keV. SN\,2018bsz is more luminous than any other non-interacting SN. In this sample, only the strongly interacting SNe of Type IIn are comparable in X-ray luminosity with SN\,2018bsz, motivating the context of interaction.
 References: This work; \citealt{fig4_1980K, fig4_1981K, fig4_1987A_1, fig4_1987A_2, fig4_1995N, fig4_1998bw, fig4_1999em, fig4_1999gi, fig4_2001ig, fig4_2002ap, fig4_2002hh, fig4_2002hi, fig4_2003bg, fig4_2004dj, fig4_2004et, fig4_2005ip, fig4_2005kd, fig4_2006bp, fig4_2006jc, fig4_2006jd, fig4_2007pk, fig4_2008ax, fig4_2008ij, fig4_2008M, fig4_2009dd, fig4_2009gj, fig4_2009mk, fig4_2010F, fig4_2010jl, fig4_2011ja, fig4_2012aw, fig4_2013ej, fig4_2017eaw}.
 }
\label{Fig:snexrays}
\end{figure}
\noindent

\begin{figure*}
\centering
 \includegraphics[width=\textwidth]{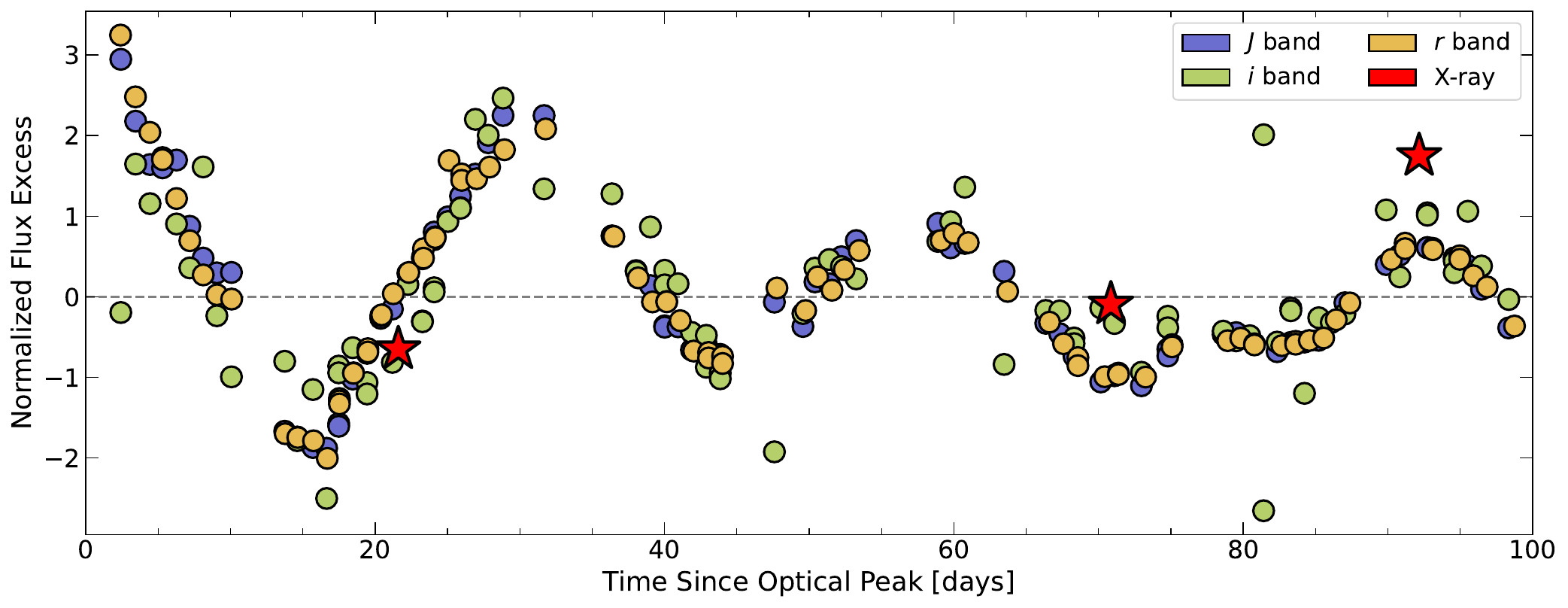}
 \caption{Renormalized flux variability for the light curve of SN\,2018bsz in the $r$, $i$, and $J$ bands, as constrained by \citet{sankar26}. Time is since optical peak (MJD 58267.5). Red stars: X-ray detections of SN\,2018bsz from this work. We note that SN\,2018bsz maintains a roughly stable level of X-ray emission well after this time window (see Figure \ref{Fig:slsnexrays}).}
\label{Fig:fluxvar}
\end{figure*}
\noindent

We place SN\,2018bsz in the context of SLSNe-I and SNe of different spectral types detected in the X-rays in Figure \ref{Fig:slsnexrays} and \ref{Fig:snexrays}, respectively. SCP\,06F6 and SN\,2018bsz bracket the entire sample of SLSNe-I of Figure \ref{Fig:slsnexrays}, further demonstrating that SCP\,06F6-like  X-ray emission reported by \citet{Levan06} is intrinsically rare, and suggesting that the lack of X-ray detection of most SLSNe is due to sensitivity limits of current instrumentation. Although these events are superluminous in the optical, this property does not translate into the X-rays. The exceptional proximity of SN\,2018bsz as one of the closest SLSNe-I discovered to date ($z = 0.0267$) played a key factor in its X-ray detection. Interestingly, Figure \ref{Fig:snexrays} shows that SN\,2018bsz is more X-ray luminous than any non-interaction powered SN. Only the strongly interacting Type IIn SNe (yellow filled circles) show comparable or larger $L_{\rm x}$, further supporting the CSM interaction interpretation of SN\,2018bsz. 

The interaction of the SN shock with dense CSM was also suggested to explain the radio emission from the SLSN-I SN\,2017ens \citep{Margutti23}. SN\,2017ens is one of only two SLSN-I detected at radio frequencies. Similarly to SN\,2018bsz, SN\,2017ens also shows the late-time emergence of hydrogen spectral features (at $\approx 190$\,d after explosion as reported by \citealt{Chen17ens}). Renormalizing the mass-loss rate inferred by \cite{Margutti23} for the same wind velocity adopted here for SN\,2018bsz, the two SLSNe-I would interestingly show comparable, large $\dot M\approx 10^{-2}\,\rm{M_{\odot}yr^{-1}}$ values, implying similarly high densities in their surroundings. However, due to the sensitivity limits of current X-ray and radio instrumentation, we note that it is possible that these densities are not representative of the SLSNe-I population and that only close SLSNe with the largest CSM densities are detectable. 

A study by \citet{sankar26} identified quasi-periodic modulations in the optical light curve of SN\,2018bsz, a phenomenology that is reminiscent of the periodic variability observed in the Type Ic SN\,2022jli \citep{chenp24}. For both SNe 2018bsz and 2022jli, the light curve modulations are observed up to few $\delta t\approx 100$\,d after optical peak, after which the light curve rapidly decays. We plot in Figure \ref{Fig:fluxvar} the normalized flux excess calculated as: 
\begin{equation}
    F_{\rm norm} = \frac{\Delta F - \mu_{\Delta F}}{\sigma_{\Delta F}} 
\end{equation}
where $\Delta F$ is the considered flux (or flux density) in a given band, $\mu_{\Delta F}$ is the mean of all flux values, and $\sigma_{\Delta F}$ is the standard deviation. Figure \ref{Fig:fluxvar} displays the resulting $r$, $i$, and $J$ band measurements derived from the analysis presented in  \citet{sankar26}, together with the X-ray measurements of SN\,2018bsz, with the goal of visualizing the observed X-ray behavior in the context of the peculiar optical evolution of the transient.

Clearly, the sparse X-ray monitoring does not allow us to draw firm conclusions regarding any potential correlation. However, the persistence of the X-ray emission in SN\,2018bsz well after the optical light curve ``shuts off'' suggests that the optical modulations and the X-ray emission are likely unrelated (SN\,2022jli was not detected in the X-rays, but it was detected in the $\gamma$-rays; \citealt{chenp24,zhang25}). 

\bigskip
\section{Summary and Conclusions}
\label{Sec:Conclusions}
We presented our multi-epoch X-ray campaign of the very nearby SLSN-I SN\,2018bsz acquired with the CXO ($\delta t \approx$ 87--303\,d rest-frame) and XRT ($\delta t \approx$ 85--866\,d rest-frame). To interpret these observations, we developed a semi-analytical framework to model the hydrodynamic expansion and thermal bremsstrahlung X-ray emission produced by the forward and reverse shocks interacting with a wind-like CSM. Our code accounts for a broken power-law ejecta profile, a wind-like CSM structure ($\rho_{\rm CSM} \propto R_{\rm sh}^{-2}$), and energy-dependent photoelectric absorption from both unshocked CSM and the Cold Dense Shell (CDS). Applied to SN\,2018bsz, our major findings can be summarized as follows:

\begin{enumerate}
    \item  With the CXO, we detect X-ray emission at the location of SN\,2018bsz peaking at $L_{\rm x} \approx 2 \times 10^{40}\,\rm{erg\, s}^{-1}$ (0.3--10\,keV) during the second epoch ($\delta t \approx$  130--142\,d), and a roughly coeval \textit{Swift}-XRT detection at $\delta t \approx$ 157\,d. A binomial significance test reveals intrinsic temporal variability at the $\approx 3\sigma$ confidence level considering the CXO data only. Although we find host-galaxy star formation may contribute to a fraction of the observed X-ray flux, the rapid X-ray variability (which is even more apparent when the CXO \emph{and} XRT data sets are considered together) makes it plausible that the signal is transient dominated.
    
    \item Under the assumption of a transient-dominated X-ray emission, we find that the overall X-ray luminosity evolution can be reasonably well-explained by thermal bremsstrahlung emission from shock interaction with the CSM, dominated at late times ($\delta t > 100$\,d) by the RS as photoelectric absorption by the CDS subsides. Comparing our shock model to the observations constrains the pre-explosion \emph{effective} mass-loss rate of the progenitor to $\dot{M} = 10^{-3}-10^{-2}\, \rm M_\odot\, yr^{-1}$ (assuming a wind velocity $v_{\rm w} = 1000\,\rm{km\,s}^{-1}$), pointing to significant stellar mass loss prior to the explosion.

    \item SN\,2018bsz is the first SLSN-I with sub-arcsecond localized X-rays. This finding is significant as the only other SLSN with claimed X-ray association (SCP\,06F6; \citealt{Levan13}) was detected with XMM, which has a substantially larger PSF. The $L_{\rm x}$ at the location of SN\,2018bsz is significantly larger than standard Type Ib/c and Type II SNe at comparable phases, reaching luminosities that overlap with the lower bound of strongly-interacting Type IIn SNe. Given that the vast majority of SLSNe-I yield X-ray non-detections the detection of X-rays at the location of SN\,2018bsz highlights the importance of its proximity ($z=0.0267$) and suggests that high surrounding densities ($\dot{M} \sim 10^{-2}\, \rm M_\odot\, yr^{-1}$, also inferred for the radio-detected SLSN-I SN\,2017ens) may not be representative of the SLSN-I progenitor population (if the detected X-rays are indeed associated with the transient).

    \item The deep, but temporally sparse X-ray follow up of SN\,2018bsz does not allow for conclusions to be drawn regarding the temporal correlation of the X-ray signal and the NIR-optical-UV quasi-periodic modulations observed during the first $\approx 100$\,d post-peak. However, the sustained X-ray emission ($>1000$\,d) beyond the fading of the optical light curve ($>100$\,d) suggests that the physical mechanism driving the optical modulations (e.g., central engine precession or inner binary interaction) is distinct from the X-ray powering source.
\end{enumerate}

In the future, continued efforts to probe the deep early time X-rays of SLSNe can assist in providing more context to these events. In the case of SN\,2018bsz, additional early time observations and denser sampling of the X-rays would have enabled constraints on the physical properties of the explosion and its surroundings, thus creating clearer distinctions between the models. Continued, deep X-ray monitoring of SN\,2018bsz over the next several years will clarify if the X-ray signal is physically related to the transient.


\vspace{5pt}\noindent
\textit{Facilities}: \textit{Swift}-XRT, CXO, XMM\\
\textit{Software}: astropy \citep{Astropy}, scipy \citep{scipy}, \texttt{CIAO} \citep{Fruscione06}

\section*{Acknowledgments}
This work was performed in part at the Aspen Center for Physics, which is supported by National Science Foundation grant PHY-2210452.
Support for this work was provided by the National Aeronautics and Space Administration through Chandra Award Number G08-19047A, issued by the Chandra X-ray Center, which is operated by the Smithsonian Astrophysical Observatory for and on behalf of the National Aeronautics Space Administration under contract NAS8-03060. This research has made use of data obtained from the Chandra Data Archive and the Chandra Source Catalog, and software provided by the Chandra X-ray Center (CXC) in the application packages CIAO and Sherpa.
This research has made use of data and software provided by the High Energy Astrophysics Science Archive Research Center (HEASARC), which is a service of the Astrophysics Science Division at NASA/GSFC.
R.M. acknowledges support by the National Science
Foundation under award No. AST-2224255, and by NASA under grant 80NSSC22K0991.
L.C. is grateful for support from NASA grants 80NSSC23K0497 and 80NSSC26K0601.
W.J.-G.\ is supported by NASA through Hubble Fellowship grant HSTHF2-51558.001-A awarded by the Space Telescope Science Institute, which is operated for NASA by the Association of Universities for Research in Astronomy, Inc., under contract NAS5-26555.
The authors also acknowledge the hospitality of the Kavli Institute for Theoretical Physics at UC Santa Barbara, where portions of this work were completed. This research was supported in part by the National Science Foundation under PHY-2309135.


\appendix

\section{Modeling Discrepancies with Ibik et al. (2025)}
\label{Sec:appendixA}
In this paper, we followed a similar procedure as \cite{ibik25} to predict the  X-ray emission from a SN RS and FS expanding into the CSM.
As part of the validation procedure for our X-ray code, we attempted to reproduce the calculation described in Appendix A of \cite{ibik25}, with an emphasis on recreating the absorption behavior shown in their Figure 16. Although we were able to recover the numerical values quoted in the text, we were unable to quantitatively reproduce the plotted absorbed X-ray light curve. \cite{ibik25} confirmed two differences between the calculation used to generate Figure 16 and the process described in the text (private communication). First, Figure 16 was generated under the assumption of a single power-law ejecta density profile ($\rho_{\rm ej} \propto R_{\rm sh}^{-n}$) rather than the broken power-law structure stated in the text. However, the numerical values quoted in the text were calculated using the broken power-law prescription as given in Equation \ref{eq:rho_ej}. Second, the expression used for $N_{\rm CSM}$ in the code of \cite{ibik25} differs from the correct equation specified in the paper (i.e., Equation \ref{eqn:ncsm}). The code implementation led 
to values of $N_{\rm CSM}$ approximately two orders of magnitude larger than those obtained from Equation \ref{eqn:ncsm}, and, in turn, to artificially suppressed observed X-ray luminosity. Both issues are corrected in the code we present and use in this paper. 

\FloatBarrier
\section{Observations}
\label{Sec:appendixB}

\begin{deluxetable}{cccccccccc}[h]
\tablecaption{Chandra X-ray Observations of SN\,2018bsz \label{Tab:18bsz_xrays}}
\tablehead{
\colhead{Start Date} & \colhead{Phase} & \colhead{Exposure} & \colhead{Net Count-Rate} & \colhead{Total Counts $^{\rm{a}}$} & \colhead{Significance} & \colhead{Photon Index} & \colhead{Abs. Flux} & \colhead{Unabs. Flux} & \colhead{Abs. Luminosity} \\
& & & (0.5--8\,keV) & (0.5--8\,keV) & (0.5--8\,keV) & & (0.3--10\,keV) & (0.3--10\,keV) & (0.3--10\,keV)\\
(MJD) & (days)& (ks) &($10^{-4}\,\rm{c\,s^{-1}}$) & & ($\sigma$) & ($\Gamma$) & ($10^{-14}\rm{erg\,s^{-1}cm^{-2}}$) & $\log_{10}(\rm{erg\,s^{-1}cm^{-2}})$ & ($10^{40}\rm{erg\,s^{-1}}$)
}
\startdata
58289.1 & 86.6 & 49.6 & $1.83 \pm 0.67$ & 11 & 3.5 & $1.58 \substack{+1.1 \\ -0.39}$ & $0.32 \substack{+0.17 \\ -0.11}$ & $-14.4 \substack{+0.22 \\ -0.17}$ & $0.47 \substack{+0.25 \\ -0.16}$ \\
58331.6 & 129.2 & 25.9 & $7.27 \pm 1.72$ & 19 & 7.9 & \multirow{2}{*}{\hspace{-5.5pt}\Big\}$\, 1.09 \substack{+0.64 \\ -0.25}$} & \multirow{2}{*}{$1.15 \substack{+0.27 \\ -0.38}$} & \multirow{2}{*}{$-13.9 \substack{+0.13 \\ -0.11}$} & \multirow{2}{*}{$1.69 \substack{+0.40 \\ -0.56}$}\\
58344.9 & 142.5 & 20.8 & $4.34 \pm 1.52$ & 9 & 3.9 & & & & \\
58505.1 & 302.6 & 49.4 & $3.37 \pm 0.91$ & 15 & $5.5 $ & $1.28 \substack{+1.2 \\ -0.22}$ & $0.49 \substack{+0.17 \\ -0.19}$ & $-14.3 \substack{+0.19 \\ -0.16}$ & $0.72 \substack{+0.25 \\ -0.27}$ \\
\enddata
\tablecomments{\textbf{a}: Total source and background counts in a $2\arcsec$ radius.}
\end{deluxetable}

\FloatBarrier
\section{X-ray Suppression in Radiative Shocks}
\label{Sec:appendixC}
\begin{figure}[H]
\centering
 \includegraphics[width=\textwidth]{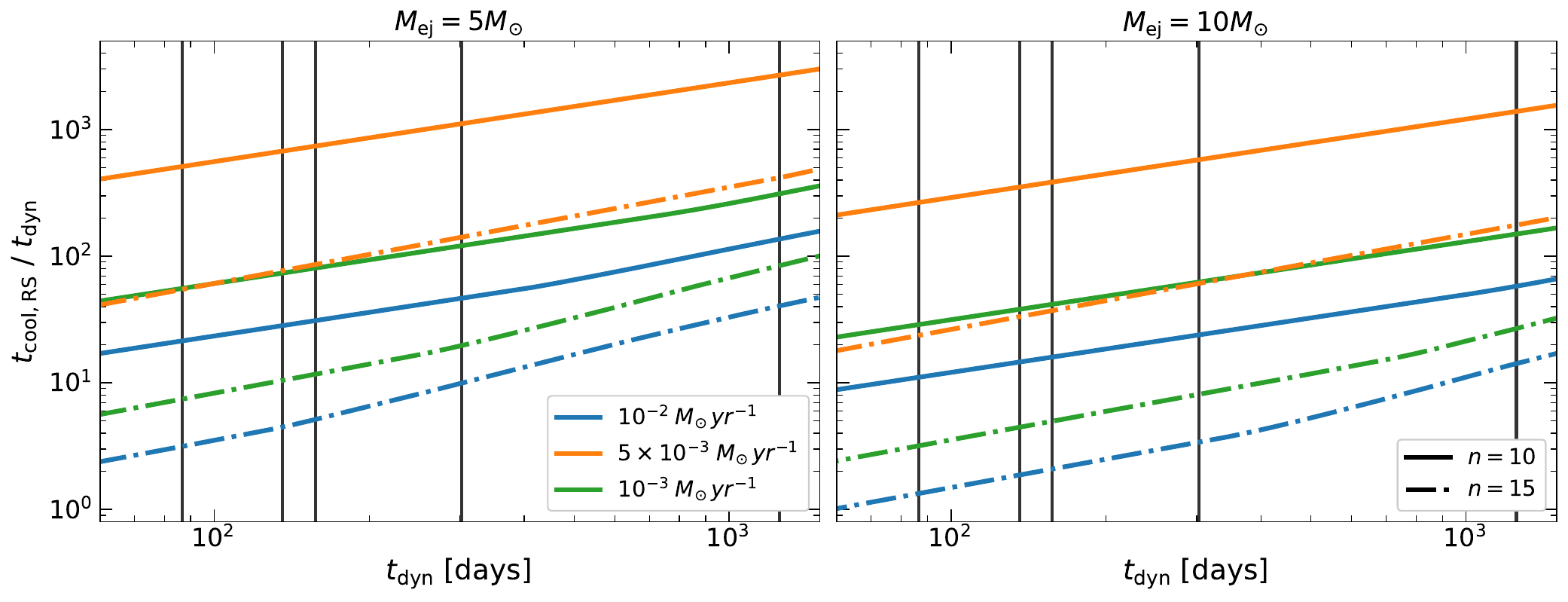}
 \caption{Visualization of the reverse shock cooling timescale over time. The vertical black lines are the times of the detections discussed in this paper. The lines in color represent the different mass-loss cases considered in our simulations, where solid lines correspond to the $n=10$ case and the dot-dashed lines are the $n=15$ case. Left: Uses parameters as in Figure \ref{Fig:5msunmodeling}. Right: Uses parameters as in Figure \ref{Fig:10msunmodeling}.}
\label{Fig:tcoolRS}
\end{figure}

When the RS is radiative, the CDS is prone to thin-shell instabilities that can lead to the suppression of X-ray emission from the hot gas (see e.g., \citealt{Metzger18, Metzger25}). This can lead to an overestimation of X-ray emission when the cooling time is short, requiring the addition of a correction factor that is not currently implemented in the code. For the case of SN\,2018bsz, as shown in Figure \ref{Fig:tcoolRS}, we find that the RS is not radiative during our times of observation. Thus, the absence of this correction does not impact our conclusions and we add a word of caution to the application of this code to cases dominated by a radiative RS.

\noindent

\FloatBarrier

\bibliographystyle{apj}
\bibliography{ref}


\shipout\box255
\end{document}